\documentclass[11pt,a4paper,english,amssymb,nofootinbib,superscriptaddress,showpacs]{revtex4}
\usepackage{lmodern}

\usepackage[T1]{fontenc}
\usepackage[latin9]{inputenc}
\usepackage{babel}
\usepackage{amsmath}
\usepackage{graphicx}
\usepackage{amssymb}
\usepackage{esint}
\usepackage[unicode=true, pdfusetitle,
 bookmarks=true,bookmarksnumbered=false,bookmarksopen=false,
 breaklinks=false,pdfborder={0 0 1},backref=false,colorlinks=false]
 {hyperref}
\def\b{\begin{equation}}
\def\e{\end{equation}}

\makeatletter
\@ifundefined{textcolor}{}
{%
 \definecolor{BLACK}{gray}{0}
 \definecolor{WHITE}{gray}{1}
 \definecolor{RED}{rgb}{1,0,0}
 \definecolor{GREEN}{rgb}{0,1,0}
 \definecolor{BLUE}{rgb}{0,0,1}
 \definecolor{CYAN}{cmyk}{1,0,0,0}
 \definecolor{MAGENTA}{cmyk}{0,1,0,0}
 \definecolor{YELLOW}{cmyk}{0,0,1,0}
 }

\usepackage{latexsym}\usepackage{bm}

\makeatother

\makeatother

\begin{document}

\title{Scattering in Topologically Massive Gravity, Chiral Gravity and the corresponding Anyon-Anyon Potential Energy}

\author{Suat Dengiz}

\email{suat.dengiz@metu.edu.tr}

\affiliation{Department of Physics,\\
 Middle East Technical University, 06800, Ankara, Turkey}

\author{Ercan Kilicarslan}

\email{kercan@metu.edu.tr}

\affiliation{Department of Physics,\\
 Middle East Technical University, 06800, Ankara, Turkey}

\author{Bayram Tekin}

\email{btekin@metu.edu.tr}

\affiliation{Department of Physics,\\
 Middle East Technical University, 06800, Ankara, Turkey}

\date{\today}

\begin{abstract}

We compute the tree-level scattering amplitude between two covariantly conserved sources in generic Cosmological Topologically Massive Gravity augmented with a
Fierz-Pauli term that has three massive degrees of freedom. We consider the Chiral Gravity limit in the anti-de Sitter space as well as the limit of Flat-Space Chiral Gravity. We show that Chiral Gravity cannot be unitarily deformed with a Fierz-Pauli mass. We calculate the non-relativistic potential energy between two point-like spinning sources. In addition to the expected mass-mass and spin-spin interactions,
there are mass-spin interactions due to the presence of the gravitational Chern-Simons term which induces spin for any massive object and turns it to an anyon.
We also show that the tree-level scattering is trivial for the Flat-Space Chiral Gravity.

\end{abstract}
\maketitle

\section{Introduction}

Gravity in $2+1$ dimensions is a fertile theoretical ground in which plethora of interesting structures and phenomena have been found \cite{DeserJackiwtHooft,DeserJackiw,Carlip_book}.
In spite of its apparent simplicity, with a vanishing Weyl tensor and no local degrees of freedom, Cosmological Einstein's theory has black hole solutions  
and the associated thermodynamics \cite{Banados}. Ultimately, of course, the goal is to find a quantum gravity theory in this lower dimensional setting and hopefully learn
about the structure of quantum gravity to build one in the more relevant $ 3+1 $-dimensions. The natural question is, after about 30 years of research, whether or not we are any closer to quantum gravity in 2+1 dimensions. 
The answer depends on one's level of optimism, while no quantum version of $ 2+1 $-dimensional Einstein's theory exists as of now, the situation is much better
when Cosmological Einstein's theory is augmented with a tuned gravitational Chern-Simons term. In this case, "Chiral Gravity" \cite{LiStrominger, Maloney} which is 
potentially a viable quantum gravity theory is conjectured to have a dual unitary boundary conformal field theory (CFT).
Generically, Einstein's Gravity plus the gravitational Chern-Simons term, that is the Topologically Massive Gravity (TMG) \cite{DJT},
has a single massive spin-2 excitation in both flat and (A)dS spacetimes. In the Chiral Gravity limit, massive graviton disappears
and one is left with a BTZ black hole with positive energy and a boundary chiral CFT from which one can relate entropy to microscopic states, 
a situation which seems to be lacking in Cosmological Einstein's theory in 2+1 dimensions.

We started this work with the following question in mind: Suppose Chiral Gravity is a viable theory, how then two covariantly 
conserved charges scatter at the lowest order with a single graviton exchange in this theory. Of course, one immediately realizes that Chiral Gravity 
cannot be easily coupled to a generic matter source but only to null matter. We then ask the more general question: How do two covariantly conserved 
particles scatter at the lowest order in Cosmological TMG augmented with a Fierz-Pauli term. Fierz-Pauli term is somewhat a headache because of its
non-covariance but at the linearized level it helps to find the propagator without worrying about gauge-fixing issues. 
And also, in 2+1 dimensions Fierz-Pauli massive gravity has a remarkable non-linear extension dubbed as "New Massive Gravity \cite{BHT} 
and even an infinite order extension of the Born-Infeld type \cite{binmg}.

Our task in this work is to find first a formal expression for the tree-level scattering in Cosmological TMG with
a Fierz-Pauli term and then consider the flat space limit. In the explicit computation of the non-relativistic potential energy,
we will recover the gravitational anyons which were found by Deser \cite{DeserAnyon} as solutions of the linearized field equations in flat space TMG.
It will be clear that because of the gravitational Chern-Simons term a point-like structureless, spinless particle with mass $m$ acts as 
if it has a spin $ \kappa m/ \mu$ where $\kappa$ is the Newton's constant and $\mu$ is the gravitational Chern-Simons parameter.
These gravitational anyons are analogs of their Abelian counterparts where any charged particle picks up
a magnetic flux when coupled to an Abelian Chern-Simons term \cite{Wilczek}.

The lay-out of the paper is as follows: We first find the particle content of the Cosmological
TMG with a Fierz-Pauli term in Section-II. That section is a generalization of the flat space version of the same theory \cite{DeserTekinMode}.
Moreover in that section we write the linearized field equations in a form which can be formally solved by the Green's function technique. 
Section III is devoted to a derivation of the tree-level scattering amplitude using the tensor decomposition of the spin-2 field in terms of its irreducible parts. 
In that section, we also give the Chiral Gravity limit of the amplitude. In section IV we limit ourselves to the flat space, where Green's functions are explicitly computable 
and the notion of potential energy makes sense as a non-relativistic approximation. There we consider two sources with mass and
spin that interact via the full TMG+ Fierz-Pauli theory and find the anyon structure of the sources. 

\section{Particle Spectrum for TMG with a Fierz-Pauli mass in (A)dS Backgrounds}

The Lagrangian density of TMG with a Fierz-Pauli mass term is 
 \begin{equation}
 {\cal L}= \sqrt{-g} \, \Bigg \{ \frac{1}{ \kappa} ( R -2 \Lambda)-\frac{m^2}{4 \kappa}(h^2_{\mu\nu}-h^2)+\frac{1}{2 \mu} \, \eta^{\mu \nu \alpha} \Gamma^\beta{_{\mu \sigma}} \Big (\partial_\nu \Gamma^\sigma{_{\alpha \beta}}+\frac{2}{3} \Gamma^\sigma{_{\nu \lambda}}  \Gamma^\lambda{_{\alpha \beta}} \Big )+ {\cal L}_{matter} \Bigg \},
\label{wtics1}
\end{equation}
where $ \kappa $ is the $2+1$-dimensional Newton's constant with mass dimension $ -1$ and $ \mu $ is a dimensionless parameter. $ \eta^{\mu \nu \alpha}$ tensor is defined in terms of the Levi-Civita symbol as $\epsilon^{\mu \nu \alpha}/\sqrt{-g}$. 
We will work with the mostly plus signature. Generically, this parity non-invariant spin-2 theory has {\it three} modes all with different masses about its flat and (A)dS backgrounds. There are constraints on the parameters 
coming from the tree-level unitarity.
Observe that when the Fierz-Pauli term vanishes, the theory reduces to TMG with a single massive spin-2 mode with $ M^2_{graviton}=\mu^2/\kappa^2+\Lambda $.
On the other hand, $ \mu \to \infty $ theory corresponds to the Fierz-Pauli massive gravity with two massive spin-2 excitations, with mass $ m $.
In the full theory, with all the parameters non-vanishing, the masses of the three modes and the unitarity regions were studied in \cite{DeserTekinMode} for {\it flat} spacetimes. Here, in this section, we generalize this result to the (A)dS backgrounds.

To find the particle spectrum about the (A)dS vacuum and the propagator, let us consider the field equations coming from the variation of (\ref{wtics1}): 
\begin{equation}
 \frac{1}{\kappa} (R_{\mu \nu}-\frac{1}{2} g_{\mu \nu} R+ \Lambda g_{\mu \nu})+\frac{1}{\mu} C_{\mu \nu}+\frac{m^2}{2 \kappa}(h_{\mu\nu}-g_{\mu\nu}h) =\tau_{\mu \nu},
\label{tmgfeq}
 \end{equation}
where $ C_{\mu \nu} $ is the symmetric, traceless and divergence-free Cotton tensor, 
\begin{equation}
 C^{\mu \nu}=\eta^{\mu \alpha \beta} \nabla_\alpha \Big ( R^\nu{_\beta}-\frac{1}{4} \delta^\nu{_\beta} R \Big ),
 \label{cotton3}
\end{equation}
which vanishes if and only if the spacetime is conformally flat. In what follows it pays to rewrite it in an explicitly symmetric form with the help of the Bianchi identity $ (\nabla_\mu G^{\mu\nu}=0) $
\begin{equation}
 C^{\mu \nu}=\frac{1}{2} \eta^{\mu \alpha \beta} \nabla_\alpha G^\nu{_\beta}+\frac{1}{2} \eta^{\nu \alpha \beta} \nabla_\alpha G^\mu{_\beta}.
 \label{coteins}
\end{equation}
 Let us consider the linearization of (\ref{tmgfeq}) about an (A)dS background $ g_{\mu\nu}=\bar{g}_{\mu\nu}+h_{\mu\nu} $ where $ \bar{R}_{\mu\nu\alpha\beta}=\Lambda (\bar{g}_{\mu \alpha} \bar{g}_{\nu \beta}-\bar{g}_{\mu \beta} \bar{g}_{\nu \alpha}) $,
which is the vacuum of (\ref{tmgfeq}) with $ m=0 $, $ \tau_{\mu\nu}=0 $:
\begin{equation}
 \frac{1}{\kappa} {\cal G}^L_{\mu \nu}+\frac{1}{2 \mu} \eta_{\mu \alpha \beta} \bar{\nabla}^\alpha {\cal G}^L_{\nu}{^\beta}+\frac{1}{2 \mu} \eta_{\nu \alpha \beta} \bar{\nabla}^\alpha {\cal G}^L_{\mu}{^\beta}+\frac{m^2}{2 \kappa}(h_{\mu\nu}-\bar{g}_{\mu\nu}h)=T_{\mu \nu}.
\label{linertmg}
\end{equation}
Here $ T_{\mu\nu}=\tau_{\mu\nu}+\Theta(h^2,h^3,...) $ and satisfies the background covariant conservation $\bar{\nabla}_\mu T^{\mu\nu}=0 $. The 2+1-dimensional linearized curvature tensors that we need are \cite{DeserTekinEnergy} 
\begin{equation}
\begin{aligned}
 {\cal G}^L_{\mu \nu}&=R^{L}_{\mu\nu}-\frac{1}{2} \bar{g}_{\mu\nu} R^{L}-2 \Lambda h_{\mu\nu}, \\
 R^{L}_{\mu\nu}&=\frac{1}{2} (\bar{\nabla}^\sigma  \bar{\nabla}_\mu h_{\nu\sigma}+\bar{\nabla}^\sigma \bar{\nabla}_\nu h_{\mu\sigma}-\bar{\square} h_{\mu\nu}-\bar{\nabla}_\mu \bar{\nabla}_\nu h ), \\
 R^{L}&=(\bar{g}^{\mu\nu}R_{\mu\nu})^{L}=-\bar{\square} h+\bar{\nabla}^\mu \bar{\nabla}^\nu h_{\mu\nu}-2 \Lambda h,
 \label{curvaturet}
 \end{aligned}
 \end{equation}
where $h=\bar{g}^{\mu\nu}h_{\mu\nu} $. To be able to identify the excitations, we have to write (\ref{linertmg}) as a source-coupled (higher derivative) wave-type equation of the form
 \begin{equation}
  (\bar{\square}-2 \Lambda-m^2_1) (\bar{\square}-2\Lambda-m^2_2) (\bar{\square}-2 \Lambda-m^2_3)h_{\mu\nu}=\tilde{T}_{\mu\nu},
 \label{wavetpemain}
 \end{equation}
where $ m_i $ correspond to the masses of the excitations. Note that in (A)dS backgrounds $ (\bar{\square}-2 \Lambda)h_{\mu\nu}=0$ is the wave equation for a massless spin-2 particle, 
hence the shifts in (\ref{wavetpemain}). Since the Levi-Civita tensor mixes various excitations in (\ref{linertmg}), we need to manipulate the equation to ``diagonalize'' it. But, before that let us note a special point in the parameter space. The divergence of (\ref{linertmg}) gives
 \begin{equation}
  m^2( \bar{\nabla}^\mu h_{\mu\nu}-\bar{\nabla}_\nu h)=0,
  \label{condtion1}
\end{equation}
yielding, for $ m^2 \neq 0 $,
 \begin{equation}
 R^L=-2 \Lambda h,\qquad  h=\frac{\kappa}{\Lambda-m^2} T, \qquad {\cal G}^L\equiv \bar{g}^{\mu\nu} {\cal G}^L_{\mu\nu} =\frac{\Lambda \kappa}{\Lambda-m^2} T. 
  \label{scaleinsstre}
 \end{equation}
 Observe that at the "partially massless point" $ m^2 = \Lambda $ and so $ h $ is not fixed and a new higher derivative gauge invariance of the form $ \delta_\xi h_{\mu\nu}=\bar{\nabla}_\mu \bar{\nabla}_\nu \xi+\Lambda \xi \bar{g}_{\mu\nu} $ appears reducing the number of the degrees of freedom by one from three to two \cite{DeserTekinMode, DeserNepomechie, DeserWaldron}. Needless to say 
 that this particular theory has no flat space limit and it shall not be considered in the rest of the paper. 
 
Now let us try to bring (\ref{linertmg}) to the form (\ref{wavetpemain}) where the masses of the excitations become apparent.
For this purpose, applying  $ \eta^{\mu \sigma \rho} \bar{\nabla}_\sigma $ to equation (\ref{linertmg}), gives 
\begin{equation}
\begin{aligned}
  \frac{1}{\kappa} \eta^{\mu \sigma \rho} \bar{\nabla}_\sigma {\cal G}^{L}_{\mu\nu}-&\frac{1}{\mu} \bar{\square}{\cal G}^{L}_\nu{^\rho}+\frac{3 \Lambda}{\mu}{\cal G}^{L}_\nu{^\rho} \\
  &+\frac{m^2}{2 \kappa} \eta^{\mu \sigma \rho} \bar{\nabla}_\sigma (h_{\mu\nu}-\bar{g}_{\mu\nu}h)= \eta^{\mu \sigma \rho} \bar{\nabla}_\sigma T_{\mu\nu}+\frac{\Lambda}{\mu} \delta^\rho{_\nu} {\cal G}^{L}
  +\frac{1}{2 \mu} \bar{\nabla}_\nu \bar{\nabla}^\rho {\cal G}^{L}-\frac{1}{2 \mu} \delta^\rho{_\nu} \bar{\square} {\cal G}^{L},
\label{ghdgh}
  \end{aligned}
  \end{equation}
where we have made use of the identity
\begin{equation}
 \begin{aligned}
\eta^{\mu \sigma \rho} \eta_{\nu \alpha \beta} = \bigg [& -\delta^\mu{_\nu} \Big ( \delta^\sigma{_\alpha} \delta^\rho{_\beta}-\delta^\sigma{_\beta} \delta^\rho{_\alpha} \Big ) 
+\delta^\mu{_\alpha} \Big ( \delta^\sigma{_\nu} \delta^\rho{_\beta}-\delta^\sigma{_\beta} \delta^\rho{_\nu} \Big ) \\
 & -\delta^\mu{_\beta} \Big ( \delta^\sigma{_\nu} \delta^\rho{_\alpha}-\delta^\sigma{_\alpha} \delta^\rho{_\nu} \Big ) \bigg ].
 \label{identity}
 \end{aligned}
\end{equation}
Using the field equation  (\ref{tmgfeq}) to eliminate the first term in (\ref{ghdgh}) one obtains
 \begin{equation}
 \begin{aligned}
 &\Big (\bar{\square}-3 \Lambda-\frac{\mu^2}{\kappa^2} \Big ) {\cal G}^{L}{_{\rho \nu}}-\frac{\mu^2 m^2}{2 \kappa^2} (h_{\rho\nu}-\bar{g}_{\rho\nu}h)-\frac{\mu m^2}{2 \kappa} \eta^{\mu \sigma}{_\rho} \bar{\nabla}_\sigma (h_{\mu\nu}-\bar{g}_{\mu\nu}h)\\
 &=\frac{\mu}{2} \, \eta_\rho{^{\mu\sigma}} \bar{\nabla}_\mu T_{\sigma\nu}+ \frac{\mu}{2} \, \eta_\nu{^{\mu\sigma}} \bar{\nabla}_\mu T_{\sigma\rho}-\frac{\mu^2}{\kappa} T_{\rho \nu} -\Lambda \bar{g}_{\rho \nu} {\cal G}^{L}-\frac{1}{2} \bar{\nabla}_\nu \bar{\nabla}_\rho {\cal G}^{L}+\frac{1}{2} \bar{g}_{\rho\nu} \bar{\square} {\cal G}^{L}.
 \label{klmapt2}
 \end{aligned}
\end{equation}
 To transform (\ref{klmapt2})
into a wave-type equation, one should write the Fierz-Pauli part in terms of the linearized Einstein tensor $ {\cal G}^L_{\mu\nu} $ and its contractions. For this purpose, let us define a new tensor
\begin{equation}
 {\cal B}_{\mu\nu} \equiv \eta_{\mu\alpha\beta} \bar{\nabla}^\alpha {\cal G}^L_{\nu}{^\beta},
 \label{redfphg}
\end{equation}
with $\bar{g}^{\mu\nu}{\cal B}_{\mu\nu}={\cal B}=0$ and $ \bar{\nabla}^\mu {\cal B}_{\mu\nu}=0 $.
Then, by plugging (\ref{redfphg}) into (\ref{klmapt2}) and hitting with $ \eta_{\rho \alpha \beta} \bar{\nabla}^\alpha $, one arrives at
\begin{equation}
\begin{aligned}
& -\frac{1}{\kappa} \eta_{\rho \alpha \beta}\bar{\nabla}^\alpha {\cal B}^\rho{_\nu}+\frac{1}{2 \mu} \eta_{\rho \alpha \beta} \eta^{\mu \sigma \rho}\bar{\nabla}^\alpha\bar{\nabla}_\sigma {\cal B}_{\mu \nu}+\frac{1}{2 \mu} \eta_{\rho \alpha \beta} \eta^{\mu \sigma \rho}\bar{\nabla}^\alpha \bar{\nabla}_\sigma {\cal B}_{\nu \mu} \\
 &+\frac{m^2}{2 \kappa}\eta_{\rho \alpha \beta} \eta^{\mu \sigma \rho} \bar{\nabla}^\alpha \bar{\nabla}_\sigma (h_{\mu\nu}-\bar{g}_{\mu\nu}h)= \eta_{\rho \alpha \beta}\eta^{\mu \sigma \rho} \bar{\nabla}^\alpha  \bar{\nabla}_\sigma T_{\mu \nu}.
\label{newdef2}
 \end{aligned}
  \end{equation}
After a somewhat lengthy but straightforward calculation, one can find the following expression
\begin{equation}
\begin{aligned}
 \frac{m^2}{2 \kappa} ( h_{\beta \nu}- \bar{g}_{\beta\nu} h)&=-\frac{m^2}{ \kappa}(\bar{\square}-2 \Lambda)^{-1} {\cal G}^L_{\beta \nu}+\frac{\Lambda}{\kappa} (\bar{\square}-2 \Lambda)^{-1} \bar{g}_{\beta \nu} {\cal G}^L \\
 &-\frac{m^2}{2 \kappa}(\bar{\square}-2 \Lambda)^{-1} \Big( \bar{g}_{\beta\nu } \bar{\square}-\bar{\nabla}_\beta \bar{\nabla}_\nu \Big)h-\Lambda (\bar{\square}-2 \Lambda)^{-1} \bar{g}_{\beta\nu} T, 
\label{explfierzpaul}
 \end{aligned}
 \end{equation}
where the inverse of an operator is a short-hand notation for the corresponding Green's function.
 By inserting (\ref{explfierzpaul}) into (\ref{klmapt2}) and using (\ref{scaleinsstre}), one finds 
   \begin{equation}
 \begin{aligned}
   &\Bigg ((\bar{\square}-3 \Lambda-\frac{\mu^2}{\kappa^2})+\frac{2 \mu^2 m^2}{ \kappa^2}(\bar{\square}-2 \Lambda)^{-1}-\frac{\mu^2 m^4}{ \kappa^2}(\bar{\square}-2 \Lambda)^{-2} \Bigg) {\cal G}^{L}{_{\rho \nu}} \\
&=\frac{\mu}{2} \, \eta_\rho{^{\mu\sigma}} \bar{\nabla}_\mu T_{\sigma\nu}+ \frac{\mu}{2} \, \eta_\nu{^{\mu\sigma}} \bar{\nabla}_\mu T_{\sigma\rho}-\frac{\mu^2}{\kappa} T_{\rho \nu}+\frac{\mu^2 m^2}{ \kappa} (\bar{\square}-2 \Lambda)^{-1} T_{\rho\nu}\\
&-\frac{\mu^2 m^2}{2 \kappa \Lambda(1-\frac{m^2}{\Lambda})}\Bigg \{(\bar{\square}-2 \Lambda)^{-1} \Big (1-m^2 (\bar{\square}-2 \Lambda)^{-1}\Big)-\frac{\kappa^2 \Lambda}{\mu^2 m^2} \Bigg \}\times \Big(\bar{g}_{\rho\nu }(\bar{\square}-2 \Lambda)-\bar{\nabla}_\rho \bar{\nabla}_\nu \Big)T,
\label{res1223}
\end{aligned}
\end{equation}
where 
\begin{equation}
 {\cal G}^L_{\rho\nu}= -\frac{1}{2} (\bar{\square}-2 \Lambda)h_{\rho\nu}+\frac{1}{2} \bar{\nabla}_\rho \bar{\nabla}_\nu h.
  \end{equation}
Equation (\ref{res1223}) is almost in the desired form, except, there is an $ h $ on the left-hand side. But this can be remedied with the help of   
$ h=\frac{\kappa}{\Lambda-m^2} T $. Then (\ref{res1223}) reads as 
\begin{equation}
{\cal O} h_{\rho\nu}=\tilde{T}_{\rho\nu},
 \end{equation}
as needed. To study the linearized gravitational degrees of freedom, let us stay away from the sources and set $T_{\rho\nu}=0$,
which gives $\tilde{T}_{\rho\nu}=0$. Then higher-order wave-type equation for Cosmological TMG with a Fierz-Pauli mass term boils down to
\begin{equation}
 \Bigg [ \Big(\bar{\square}-3 \Lambda-\frac{\mu^2}{\kappa^2} \Big )(\bar{\square}-2 \Lambda)^2+\frac{2 \mu^2 m^2}{ \kappa^2}(\bar{\square}-2 \Lambda)-\frac{\mu^2 m^4}{ \kappa^2} \Bigg]h_{\rho \nu}=0,
\label{vacuum3}
 \end{equation}
which, in flat space, reduces to the known form \cite{DeserTekinMode}
 \begin{equation}
 \Bigg [(\partial^2)^3-\frac{\mu^2}{\kappa^2}(\partial^2)^2+\frac{2 \mu^2 m^2}{ \kappa^2}\partial^2 -\frac{\mu^2 m^4}{ \kappa^2} \Bigg ] h_{\rho \nu}=0,
 \label{massflat}
\end{equation}
that has three distinct real roots corresponding to the masses of the excitations only if $ \mu^2/m^2 \kappa^2 \geq 27/4 $. 

To read the masses of the excitations in the (A)dS background, equation (\ref{vacuum3}) rewritten as (\ref{wavetpemain}) with a vanishing right-hand side,
yields a cubic equation 
\begin{equation}
 M^6-(\Lambda+\frac{\mu^2}{\kappa^2})M^4+\frac{2\mu^2 m^2}{\kappa^2}M^2-\frac{\mu^2 m^4}{\kappa^2}=0,
 \label{adsmassspec}
\end{equation}
where again the $3 $ roots $ M_i $ are the masses of the excitations. In general there are complex roots unless
\begin{equation}
 1+\frac{9\kappa^2\Lambda}{\mu^2}(1-\frac{3m^2}{4\Lambda}) \geq \frac{\Lambda}{m^2}(1+\frac{\kappa^2\Lambda}{\mu^2})^2,
\end{equation}
which guarantees the non-negativity of the discriminant. The explicit forms of all the 3 roots are somewhat cumbersome to write, therefore we shall not display them here.
But let us note the following cases:
\begin{enumerate}
 \item When $ \frac{\mu^2}{\kappa^2} =-\Lambda $, which is the Chiral Gravity limit, there are two tachyonic excitations unless $ m=0 $.
 This means that Chiral Gravity cannot be deformed unitarily with a Fierz-Pauli mass.
 \item In the $m^2=0$ limit caution must be exercised: Namely there are ostensibly two solutions $ M^2=0 $ and $ m^2=\Lambda+\frac{\mu^2}{\kappa^2} $.
 The latter is the well-known massive mode while the first solution does not actually exist. This can be seen from the equations if one 
 started with $ m^2=0 $ in the beginning \cite{Carlip, GursesSismanTekin} .
 \item As a specific example, let us take $ m^2=\frac{8\Lambda}{9} $, and $\frac{\mu^2}{\kappa^2}=3 \Lambda $ then
 all the 3 roots are equal 
 \begin{equation}
   M^2_1= M^2_2= M^2_3=\frac{4\Lambda}{3},
 \end{equation}
which obey the Higuchi bound \cite{higuchi} in dS space $ (M^2 > \Lambda > 0) $ but do not obey the Breitenlohner-Freedman bound \cite{bf} in AdS $ ( M^2 > \Lambda ) $.
At this specific point, the fact that helicity+2 and helicity-2 modes have the same mass in dS does not show that parity symmetry is restored.  
 \item In the $ \mu \to \infty $ limit, which is the Einstein-Fierz-Pauli theory one gets two excitations with the same mass $ m $.
 \end{enumerate}

 Having found the spectrum of the full theory, we now move on to calculate the tree-level scattering amplitude between two conserved currents $ (\bar{\nabla}_\mu T^{\mu\nu}=0). $

\section{Scattering Amplitude in generic TMG Plus Fierz-Pauli Theory}

Equation (\ref{res1223}) is of the form $ {\cal O} h_{\mu\nu}=\tilde{T}_{\mu\nu} $, where $ {\cal O}$ is a complicated operator.
Since not every component of $h_{\mu\nu}$ is dynamical, it pays to decompose the spin-2 field in terms of transverse helicity-2 $ ( h^{TT}_{\mu\nu}) $,
helicity-1  $(V^\mu) $ and helicity-0 components $(\phi, \psi)$ as
\begin{equation}
  h_{\mu\nu} \equiv  h^{TT}_{\mu\nu}+\bar{\nabla}_{(\mu}V_{\nu)}+\bar{\nabla}_\mu \bar{\nabla}_\nu \phi+\bar{g}_{\mu\nu} \psi.
  \label{dectth}
 \end{equation}
 Taking the trace of (\ref{dectth}) and the divergence of (\ref{condtion1}) lead to an elimination of $\phi $ and a relation between $ h $ and $ \psi $ as
\begin{equation}
 h=\frac{1}{\Lambda} (\bar{\square}+3 \Lambda) \psi.
\end{equation}
Then the trace of the field
 equations (\ref{scaleinsstre}) yield
 \begin{equation}
  \psi = \frac{\kappa}{1-\frac{m^2}{\Lambda}} (\bar{\square}+3 \Lambda)^{-1} T.
  \label{psittt}
 \end{equation}
 To relate $  h^{TT}_{\mu\nu} $ to the source, one needs to use the Lichnerowicz operator, $ \triangle^{(2)}_L $ acting on the symmetric spin-2 tensors as 
  \begin{equation}
  \triangle^{(2)}_L h_{\mu\nu} =-\bar{\square} h_{\mu\nu}-2 \bar{R}_{\mu\rho\nu\sigma} h^{\rho\sigma}+2\bar{R}^\rho{_{(\mu}} h_ {\nu)\rho},
 \end{equation}
with the following properties \cite{Porrati} 
 \begin{equation}
  \begin{aligned}
    \triangle^{(2)}_L \nabla_{(\mu}V_{\nu)}&= \nabla_{(\mu} \triangle^{(1)}_L V_{\nu)} , \qquad  \quad \,\,\, \triangle^{(1)}_L V_{\mu}=(-\square+\Lambda)V_\mu, \\
    \nabla^\mu \triangle^{(2)}_L h_{\mu\nu} &=\triangle^{(1)}_L \nabla^\mu h_{\mu\nu}, \qquad  \,\, \nabla^\mu \triangle^{(1)}_L V_{\mu}= \triangle^{(0)}_L \nabla^\mu V_\mu, \\
    \triangle^{(2)}_L g_{\mu\nu} \phi &= g_{\mu\nu}\triangle^{(0)}_L \phi, \qquad  \qquad \quad \triangle^{(0)}_L \phi=-\square \phi.
  \label{lichng}
  \end{aligned}
 \end{equation}
 Then, with the help of (\ref{lichng}), transverse-traceless part of the linearized Einstein tensor 
 ${\cal G}^{TT}_L{_{\rho \nu}}$ can be written in terms of Lichnerowicz operator as 
 \begin{equation}
  {\cal G}^{TT}_L{_{\rho \nu}}=\frac{1}{2} \Big ( \triangle^{(2)}_L-4\Lambda \Big ) h^{TT}_{\rho\nu}.
  \label{einstett2}
 \end{equation}
Substituting (\ref{einstett2}) into (\ref{res1223}) gives a relation between the transverse-traceless field and the sources 
 \begin{equation}
  \begin{aligned}
    h^{TT}_{\rho\nu}&=\mu \, {\cal O}^{-1} (\bar{\square}-2 \Lambda)^{2} \eta_\rho{^{\mu\sigma}} \bar{\nabla}_\mu T^{TT}_{\sigma\nu}+ \mu \,{\cal O}^{-1} (\bar{\square}-2 \Lambda)^{2} \eta_\nu{^{\mu\sigma}} \bar{\nabla}_\mu T^{TT}_{\sigma\rho} \\
&-\frac{2 \mu^2}{\kappa} {\cal O}^{-1} (\bar{\square}-2 \Lambda)^{2}T^{TT}_{\rho \nu}+\frac{2 \mu^2 m^2}{ \kappa} {\cal O}^{-1} (\bar{\square}-2 \Lambda) T^{TT}_{\rho\nu},
 \label{htt}
 \end{aligned}
 \end{equation}
where the scalar Green's function $ {\cal O}^{-1} $ is
\begin{equation}
  {\cal O}^{-1} \equiv \Bigg \{ \Big [ (\bar{\square}-2 \Lambda)^{2} \Big (\bar{\square}-3 \Lambda-\frac{\mu^2}{\kappa^2} \Big)+\frac{2 \mu^2 m^2}{ \kappa^2}(\bar{\square}-2 \Lambda)-\frac{\mu^2 m^4}{ \kappa^2}\Big ] \times  \Big ( \triangle^{(2)}_L-4\Lambda \Big ) \Bigg \}^{-1}. 
\end{equation}
Additionally, doing a similar tensor decomposition (\ref{dectth}) for $ T_{\rho\nu} $, one can write the transverse-traceless part $T^{TT}_{\rho\nu} $ \cite{Gullu:2009vy} as 
\begin{equation}
 T^{TT}_{\rho\nu} \equiv T_{\rho\nu}-\frac{1}{2} \bar{g}_{\rho\nu}T+\frac{1}{2} \Big (\bar{\nabla}_\rho \bar{\nabla}_\nu + \Lambda \bar{g}_{\rho\nu} \Big ) \times (\bar{\square}+3 \Lambda )^{-1} T.
\label{ttstrener}
 \end{equation}
With all the above results, we are ready to write the tree-level scattering amplitude between two sources as
\begin{equation}
\begin{aligned}
 {\cal A}&=\frac{1}{4} \int d^3 x \sqrt{-\bar{g}} T^{'}_{\rho\nu}(x) h^{\rho\nu}(x) \\
 &=\frac{1}{4} \int d^3 x \sqrt{-\bar{g}} (T^{'}_{\rho\nu} h^{TT\rho\nu}+T^{'} \psi ).
 \label{scatdef}
\end{aligned}
 \end{equation}
 Finally plugging (\ref{psittt}), (\ref{einstett2}) and (\ref{ttstrener}) into (\ref{scatdef}), one gets the amplitude as
\begin{equation}
 \begin{aligned}
  4{\cal A}&=2\mu T^{'}_{\rho\nu} {\cal O}^{-1} (\bar{\square}-2 \Lambda)^{2} \eta^{\rho\mu\sigma}\bar{\nabla}_\mu T_\sigma{^\nu}-\frac{2 \mu^2}{\kappa}T^{'}_{\rho\nu} {\cal O}^{-1}(\bar{\square}-2 \Lambda)(\bar{\square}-2 \Lambda-m^2) T^{\rho\nu} \\
  &-\frac{\mu^2}{\kappa}T^{'}_{\rho\nu} {\cal O}^{-1} (\bar{\square}-2 \Lambda)(\bar{\square}-2 \Lambda-m^2) (\bar{\nabla}^\rho \bar{\nabla}^\nu + \Lambda \bar{g}^{\rho\nu}) \times \Big (\bar{\square}+3 \Lambda \Big )^{-1} T \\
&+\frac{\mu^2}{\kappa} T^{'} {\cal O}^{-1}(\bar{\square}-2 \Lambda)(\bar{\square}-2 \Lambda-m^2) T + \frac{\kappa}{1-\frac{m^2}{\Lambda}} T^{'} (\bar{\square}+3 \Lambda)^{-1} T,
\label{mainressct}
\end{aligned}
\end{equation}
where for notational simplicity we have suppressed the integral signs.

The pole structure of the full theory is highly complicated: there are apparently 4 poles. But in the most general case, it is hard to see from the scattering amplitude,
whether the fourth pole, besides the 3 which are exactly the roots of the cubic equation (\ref{adsmassspec}), is unphysical or not. In any case, the introduction of the Fierz-Pauli
term served its purpose of making the propagator invertible and hence we can now set it to zero and consider the most promising limit of the general theory that is
the Chiral Gravity with $ m^2=0, \, \mu^2/\kappa^2=-\Lambda $. Strictly speaking one must keep $ h=0 $, to get the Chiral Gravity. Hence, we must also set $ T=0 $. 
Then the amplitude in Chiral gravity reads   
\begin{equation}
 4{\cal A}=2 \mu T^{'}_{\rho\nu} \Bigg \{\Big (\bar{\square}-2 \Lambda \Big) \times  \Big (\triangle^{(2)}_L-4\Lambda \Big ) \Bigg \}^{-1} \Bigg (\eta^{\rho\mu\sigma}\bar{\nabla}_\mu T_\sigma{^\nu}-\frac{\mu}{\kappa}T^{\rho\nu} \Bigg),
\end{equation}

recall that the theory is valid in AdS with $ \Lambda <0 $. As expected the massive mode dropped out.

\section{Flat Space Considerations}

In this section, we will study the flat space limit of the scattering amplitude (\ref{mainressct}) in various theories 
and the corresponding Newtonian potential energy ($ {\cal U} $) between two localized conserved spinning point-like sources defined by 
\begin{equation}
 T_{00}= m_a \delta^{(2)}({\bf x}-{\bf x}_a), \quad T^{i}{_0}=- \frac{1}{2} J_a \epsilon^{ij} \partial_j \delta^{(2)}({\bf x}-{\bf x}_a),
 \label{sourcegen}
\end{equation}
where $ a=1,2 $ and $ m_a $ and $ J_a $ refer to the mass and spin respectively. (Note that  spin in $2+1$ dimensions is a pseudoscalar quantity which can be negative or positive.)

\subsection{Scattering of Anyons in TMG with a Fierz-Pauli term}
 In the $ \Lambda \to 0 $ limit,  (\ref{mainressct}) gives the tree-level scattering amplitude in flat space for the TMG with a Fierz-Pauli mass:
 \begin{equation}
 \begin{aligned}
  4{\cal A}&=-2 \mu T^{'}_{\rho\nu} \frac{\partial^2}{\partial^4 (\partial^2-\frac{\mu^2}{\kappa^2})+\frac{2 \mu^2 m^2}{\kappa^2} \partial^2-\frac{\mu^2 m^4}{\kappa^2}} \eta^{\rho\mu\sigma}\partial_\mu T_\sigma{^\nu}\\
  &+\frac{2 \mu^2}{\kappa}T^{'}_{\rho\nu}  \frac{\partial^2-m^2}{\partial^4 (\partial^2-\frac{\mu^2}{\kappa^2})+\frac{2 \mu^2 m^2}{\kappa^2} \partial^2-\frac{\mu^2 m^4}{\kappa^2}} T^{\rho\nu}
   -\frac{\mu^2}{\kappa} T^{'}\frac{\partial^2-m^2}{\partial^4 (\partial^2-\frac{\mu^2}{\kappa^2})+\frac{2 \mu^2 m^2}{\kappa^2} \partial^2-\frac{\mu^2 m^4}{\kappa^2}} T. 
\label{scatflat2}
  \end{aligned}
\end{equation}
As long as one is looking at the generic theory where the masses are distinct, one can fractionally decompose the propagator as  
\begin{equation}
\frac{\partial^2-m^2}{\partial^4 (\partial^2-\frac{\mu^2}{\kappa^2})+\frac{2 \mu^2 m^2}{\kappa^2} \partial^2-\frac{\mu^2 m^4}{\kappa^2}} 
\equiv \sum_{k=1}^{3} \prod_{\substack{r=1 \\ r \neq k}}^3 \frac{(M^2_k-m^2)}{(M^2_k-M^2_r)} G_k({\bf x}, {\bf x}^{'},t,t^{'}),
 \label{decgreen1}
  \end{equation}
  where the scalar Green's function is $  G_k({\bf x}, {\bf x}^{'},t,t^{'})= (\partial^2-M^2_k)^{-1} $
 and $M_k =M_k (\kappa^2, \mu^2,m^2) $, $ k=1,2,3 $, are the generic roots of the equation (\ref{adsmassspec}). 
Substituting (\ref{decgreen1}) into (\ref{scatflat2}) and using (\ref{sourcegen}) and carrying out the time integrals yield  
\begin{equation}
 \begin{aligned}
4\,{\cal U}&=\sum_{k=1}^{3} \prod_{\substack{r=1 \\ k \neq r}}^3(M^2_k-M^2_r)^{-1} \Bigg \{\frac{\mu^2 M^2_k}{\kappa} \Big (\frac{\kappa m_2 }{\mu} J_1 +\frac{\kappa m_1 }{\mu} J_2  + J_1 J_2(1-\frac{ m^2 }{M^2_k}) \Big ) \\
& \qquad \qquad\qquad\qquad\qquad \times \int d^2 x \int d^2 x^{'}\,\,\,\delta^{(2)}({\bf x}^{'}-{\bf x}_2) \partial_i \partial^i \hat{G}_k({\bf x}, {\bf x}^{'}) \delta^{(2)}({\bf x}-{\bf x}_1) \\
&\qquad \qquad\qquad\qquad\qquad +  \frac{\mu^2 m_1 m_2 }{\kappa} ( M^2_k-m^2) \int d^2 x \int d^2 x^{'}\,\,\, \delta^{(2)}({\bf x}^{'}-{\bf x}_2) \hat{G}_k({\bf x}, {\bf x}^{'}) \delta^{(2)}({\bf x}-{\bf x}_1) \Bigg \},
  \end{aligned}
\end{equation}
where the potential energy is defined as $ {\cal U}={\cal A}/t $  (See  \cite{gullu_spin}) and the time-integrated Green's function reads
\begin{equation}
  \hat{G}_k ({\bf x},{\bf x}^{'})=\int d t^{'} \, G_k ({\bf x},{\bf x}^{'},t,t^{'})=\frac{1}{2 \pi}  K_{0}\, (M_k \lvert {\bf x}-{\bf x}^{'} \rvert).
\end{equation}
 Finally using the recurrence relation between the modified Bessel functions 
\begin{equation}
  \vec{\nabla}^2 K_0 (M_k r)= \frac{M^2_k}{2} \Big (K_0 (M_k r)+K_2 (M_k r) \Big ),
\label{recur}
\end{equation}
where $ r = \lvert {\bf x}_1-{\bf x}_2 \rvert$, one obtains
\begin{equation}
 \begin{aligned}
{\cal U}&=\sum_{k=1}^{3} \prod_{\substack{r=1 \\ k \neq r}}^3(M^2_k-M^2_r)^{-1} \Bigg \{\frac{\mu^2 M^4_k}{16 \pi \kappa} \Big (J^{tot}_1 J^{tot}_2-\frac{\kappa^2 m_1 m_2}{\mu^2}- \frac{ m^2 J_1 J_2 }{M^2_k} \Big )  K_{2}(M_k r) \\
&\qquad \qquad\qquad\qquad\qquad +  \frac{\mu^2 M^4_k}{16 \pi \kappa}\Big [\frac{2 m_1 m_2 }{ M^2_k}(1-\frac{m^2 }{M^2_k} )+ \Big (J^{tot}_1 J^{tot}_2-\frac{ \kappa^2 m_1 m_2}{\mu^2}- \frac{ m^2 J_1 J_2 }{M^2_k} \Big )\Big ] \\
&\qquad \qquad\qquad\qquad\qquad\times K_{0}\, (M_k r) \Bigg \}.
  \label{newtpot1}
  \end{aligned}
\end{equation}
 We defined the total spin as the original spin of the source plus the induced spin due to the gravitational Chern-Simons term, turning the source to an anyon  \cite{DeserAnyon}: 
\begin{equation}
 J^{tot}_a \equiv J_a+ \frac{\kappa m_a}{\mu}, \quad a=1,2.
\end{equation}
Our result not only reveals the anyon structure of the sources, but it also describes how anyons scatter at small energies. For other works on gravitational anyons see \cite{DeserMcCarthy, Ortiz, Clement, EderyAnyon}.
Observe that  depending on the choice of $(J_a, m_a, m^2)$, $ {\cal U} $ can be either negative or positive or even it could vanish. 

Let us now consider the short and large distance limits of the anyon-anyon potential energy. First, in short distances, since the Bessel function behave as
\begin{equation}
 K_0 (M_k r) \sim -\ln(M_k r)-\gamma_{E}, \quad   K_2 (M_k r) \sim \frac{2}{M^2_k} \frac{1}{r^2} , 
\label{limitc}
 \end{equation}
the potential energy reads
  \begin{equation}
 \begin{aligned}
{\cal U}& \sim \sum_{k=1}^{3} \prod_{\substack{r=1 \\ k \neq r}}^3(M^2_k-M^2_r)^{-1} \Bigg \{\frac{\mu^2 M^2_k}{8 \pi \kappa} \Big (J^{tot}_1 J^{tot}_2-\frac{\kappa^2 m_1 m_2}{\mu^2}- \frac{ m^2 J_1 J_2 }{M^2_k} \Big ) \times \frac{1}{r^2} \\
&\qquad \qquad\qquad\qquad\qquad - \frac{\mu^2 M^4_k}{16 \pi \kappa}\Big [\frac{2 m_1 m_2 }{ M^2_k}(1-\frac{m^2 }{M^2_k} )+ \Big (J^{tot}_1 J^{tot}_2-\frac{\kappa^2 m_1 m_2}{\mu^2}- \frac{ m^2 J_1 J_2 }{M^2_k} \Big )\Big ] \\
&\qquad \qquad\qquad\qquad\qquad\times \Big (\ln(M_k r)+\gamma_{E} \Big) \Bigg \},
  \label{newtpot3}
  \end{aligned}
\end{equation}
where $ \gamma_E $ is the Euler-Mascheroni constant. On the other side, for large distances, since the Bessel functions decay as 
\begin{equation}
  K_n(M_k r) \sim \sqrt{\frac{\pi}{ 2 M_k r}} \, e^{-M_k r},
 \end{equation}
the potential energy becomes
\begin{equation}
 \begin{aligned}
{\cal U}& \sim \sum_{k=1}^{3} \prod_{\substack{r=1 \\ k \neq r}}^3(M^2_k-M^2_r)^{-1} \frac{\mu^2 M^4_k}{8 \pi \kappa}\Big [\frac{ m_1 m_2 }{ M^2_k}(1-\frac{m^2 }{M^2_k} )+ \Big (J^{tot}_1 J^{tot}_2-\frac{\kappa^2 m_1 m_2}{\mu^2}- \frac{ m^2 J_1 J_2 }{M^2_k} \Big )\Big ] \\
&\qquad \qquad\qquad\qquad\qquad\qquad\quad\times \sqrt{\frac{\pi}{ 2 M_k r}} \, e^{-M_k r}.
 \label{newtpot5}
\end{aligned}
\end{equation}
  
\subsection{Scattering of Anyons in TMG}

 We now consider the scattering of anyons and find the related potential energy for TMG without the Fierz-Pauli mass. 
 Taking $ m^2 \to 0 $ and $ \Lambda \to 0 $ limits in (\ref{mainressct}) in that order to avoid the van-Dam Veltman-Zakharov discontinuity (vDVZ) \cite{Porrati, Kogan}, one arrives at 
\begin{equation}
 \begin{aligned}
  4{\cal A}&=-2\mu T^{'}_{\rho\nu} \frac{1}{(\partial^2-\frac{\mu^2}{\kappa^2})\partial^2} \eta^{\rho\mu\sigma}\partial_\mu T_\sigma{^\nu}+\frac{2 \mu^2}{\kappa}T^{'}_{\rho\nu} \frac{1}{(\partial^2-\frac{\mu^2}{\kappa^2})\partial^2}   T^{\rho\nu} \\
  &-\frac{\mu^2}{\kappa} T^{'} \frac{1}{(\partial^2-\frac{\mu^2}{\kappa^2})\partial^2} T + \kappa T^{'} \frac{1}{\partial^2} T,
\label{anyon-anyon1}
\end{aligned}
\end{equation}
 which generically has a massive and a massless modes. The explicit computation of the potential energy follows along the same line of the previous section. 
 One finally arrives at the anyon-anyon scattering potential energy in TMG:
\begin{equation}
 {\cal U}= \frac{\kappa m^2_g}{16 \pi }\bigg \{ \Big ( J^{tot}_1 J^{tot}_2-\frac{m_1 m_2}{m^2_g} \Big )K_2(m_g r)+ \Big ( J^{tot}_1 J^{tot}_2+\frac{m_1 m_2}{m^2_g} \Big )K_0(m_g r ) \bigg \},
\label{tmganyan}
 \end{equation}
where $m^2_g= \mu^2/\kappa^2 $.

Let us now check the small and large distance behaviors of the potential energy: First of all, for small separations, one obtains  
\begin{equation}
\begin{aligned}
 {\cal U}  \sim \frac{\frac{\kappa}{8 \pi}  \Big ( J^{tot}_1  J^{tot}_2-\frac{m_1 m_2}{m^2_g} \Big )}{ r^2}  
-\frac{\kappa m^2_g}{16 \pi}  \Big ( J^{tot}_1 J^{tot}_2+\frac{m_1 m_2}{m^2_g} \Big ) \Big (\ln(m_g r)+\gamma_{E} \Big ).
  \end{aligned}
\end{equation}
At large distances (\ref{tmganyan}) behaves as 
 \begin{equation}
 {\cal U} \sim \frac{\kappa m^2_g J^{tot}_1 J^{tot}_2}{8  \pi} \sqrt{\frac{\pi}{ 2 m_g r}} \, e^{-m_g r},
\end{equation}
which of course could be repulsive or attractive. For the specific case of the tuned spin $ J = - \kappa m/ \mu$, there is no interaction at large separations.

\subsection{Scattering of Anyons in Flat-Space Chiral Gravity}

In \cite{Bagchi} as an example of the holographic correspondence between a gravitational theory in flat space and a conformal field theory (CFT) in a lower dimensional space (which is akin
to the AdS/CFT correspondence) a chiral gravity is constructed as a limit of TMG, which the authors dubbed ``Flat-Space Chiral Gravity'' and 
showed that a pure gravitational Chern-Simons term with level $k$, i.e., 
\begin{equation}
 S=\frac{k}{4 \pi} \int d^3 x \sqrt{-g} \, \eta^{\mu \nu \alpha} \Gamma^\beta{_{\mu \sigma}} \Big (\partial_\nu \Gamma^\sigma{_{\alpha \beta}}+\frac{2}{3} \Gamma^\sigma{_{\nu \lambda}}  \Gamma^\lambda{_{\alpha \beta}} \Big ),
\end{equation}
is dual to a CFT with a chiral charge $ c=24 $.

Here we consider the scattering amplitude and the Newtonian potential energy in Flat-Space Chiral Gravity. To do so, let us note how Flat-Space Chiral Gravity arises from TMG:
\begin{equation}
 \kappa \to \infty , \quad \mu \to \frac{2 \pi}{k}.
\end{equation}
From (\ref{mainressct}) and with $ T=0 $ and $ m^2=0 $, one obtains
 \begin{equation}
 4{\cal A}=-\frac{4 \pi}{k} T^{'}_{\rho\nu} \frac{1}{\partial^4} \eta^{\rho\mu\sigma} \partial_\mu T_\sigma{^\nu}. 
\label{flatspace}
 \end{equation}
 We need to construct a covariantly conserved traceless source
 \footnote{After time integration of $ (\partial^4)^{-1} $, one gets the biharmonic Green's function:
 $\int dt (\partial^4)^{-1} ({\bf x}, {\bf x}^{'},t,t^{'})= -\frac{\lvert {\bf x}-{\bf x}^{'} \rvert^2}{8 \pi} \Big ( \ln \frac{\lvert {\bf x}-{\bf x}^{'} \rvert}{\zeta} -1 \Big ). $}. To do this we can write the Minkowski space in null coordinates as $ ds^2=-dudv+dy^2 $, with $ u=t+x $ and
 $ v=t-x $. Then the vector $ l_\mu \equiv \partial_\mu u $ satisfies $ l^\mu =-\delta^\mu_v $ and $ l_\mu l^\mu=0 $.
 Therefore the null source should read $ T^{\mu\nu} \sim l^\mu l^\nu  $. Together with the condition $ \nabla_\mu T^{\mu\nu}=0 $,
 we have $ T^{\mu\nu}=E \delta(u) \delta(y) \delta^\mu_v \delta^\nu_v $. Substitution of this in (\ref{flatspace}) yields a trivial scattering amplitude.

\section{conclusion}

We have studied the 2+1-dimensional Cosmological Topologically Massive gravity (CTMG) augmented by a Fierz-Pauli mass term in (A)dS and flat backgrounds in detail. 
We first found the particle spectrum of the full theory in (A)dS and the computed the tree-level scattering amplitude between two conserved energy-momentum tensors.
We also looked at the Chiral Gravity limit we found that Chiral Gravity cannot be unitarily deformed with Fierz-Pauli mass term. In flat background, we studied the potential energy between two point-like spinning sources and obtained the anyon nature of the sources
that arises due to the gravitational Chern-Simons term. In the Flat-Space Chiral Gravity, scattering at the tree-level is trivial.

\section{\label{ackno} Acknowledgments}

We would like to thank T. C. Sisman and I. Gullu for useful discussions. The work of B.T. and S.D. is supported by the TUBITAK Grant No.113F155. E.K. is supported by the TUBITAK PhD Scholarship.

\end{document}